\documentclass[aps,preprint,prd, showpacs, showkeys]{revtex4-1}
\usepackage{dcolumn}
\usepackage{epsfig,amssymb,multirow,amsmath, bm,xcolor,soul,slashed}
\usepackage{graphicx}

\begin{document}
\title{On the role of the Vacuum Energy in the Thermodynamics of Neutron Matter }

\author{J.P.W. Diener$^{a}$ and F.G. Scholtz$^{b,c}$}
 \affiliation{}
 \affiliation{$^{a}$Botswana International University of Science and Technology (BIUST), Palapye, Botswana\\$^{b}$National Institute for Theoretical Physics (NITheP), Stellenbosch 7600, South Africa\\ $^{c}$Institute of Theoretical Physics, University of Stellenbosch, Stellenbosch 7600, South Africa}

\date{\today}

\begin{abstract}
The only way neutron matter can couple to the electromagnetic field is through an anomalous coupling, which plays an important role in the thermodynamics of pure neutron matter. Such theories are, however, perturbatively non-renormalisable, which presents a difficulty in terms of the unambiguous treatment of the divergencies.  Here we show that despite this, an unambiguous expression can be obtained for the vacuum energy contribution to the grand canonical potential in the case of a constant magnetic field.  We find that this contribution is quite small, which justifies the no-sea approximation usually made.  We also discuss the density and temperature dependence of the full grand canonical potential.   

\end{abstract}

\maketitle 

\section{Introduction}

Central to our understanding of neutron stars and core-collapse supernovae is the nuclear matter equation of state \cite{Bethe,Lattimer1}. The nuclear matter equation of state shows little variation up to saturation density ($\rho_s=0.16$ fm$^{-3}$), whether the matter is magnetised or not \cite{Bethe,Lattimer1,Diener1}, but there are many outstanding issues at the densities and magnetic field strengths predicted to be found in neutron stars.  Typically, these densities are well above saturation density while the magnetic field can reach up to  $10^{18}$G in the star's core \cite{KandK, FandR}.  The analysis of these systems therefore requires knowledge of the equation of state at high densities and in the presence of strong magnetic fields.   

Magnetised dense nuclear matter has already been studied for several decades. The approach of \cite{Thor1,Thor2} was based on non-relativistic Thomas-Fermi theory 
and mainly focused on the  neutron star crust, while relativistic studies used the MIT bag model \cite{Chak1}, relativistic Hartree theory \cite{Chak2} and relativistic mean-field theories \cite{Zhan}.  The relativistic mean-field computations were further refined in \cite{Lattimer2}  with emphasis on the roles of Landau quantisation and the anomalous magnetic dipole moment of the nucleons.  The latter was taken into account through an anomalous coupling term.  More recently these issues were further explored in \cite{Ferr1,Strict}, which emphasised the pressure anisotropy resulting from the breaking of rotational symmetry by the magnetic field (see also \cite{Pot} and \cite{Ferr2}).  A detailed analysis of the effect of the anomalous coupling was also carried out in \cite{Ferr3} where it was concluded that it plays an insignificant role for protons, but that it can be important for neutrons as it is the only coupling mechanism to the magnetic field.  

These computations all shared the ``no-sea'' approximation, i.e. the divergent contribution of the vacuum energy to the grand canonical potential was ignored.  The main reason for this is that the anomalous coupling renders the theory non-renormalisable, which makes consistent control of the ultra-violet divergencies difficult.  A minimal subtraction scheme was proposed in \cite{Diener2}, while a more recent attempt to include vacuum energy effects was made in \cite{Ferr4} by regulating the divergent vacuum energy contribution through a momentum cut-off of the order of the chemical potential.  Although these approaches are physically sensible, they remain somewhat ad hoc and one would like to find a more rigorous way of computing the contribution of the vacuum energy to the grand canonical potential.  This is the aim of this paper, where we argue that despite the non-renormalisability of the theory, it is still possible to unambiguously compute the contribution of the vacuum energy to the grand canonical potential.  Here we demonstrate this in the case of pure neutron matter in a magnetic field where the anomalous coupling is most relevant.

This paper is organised as follows: Section \ref{one} briefly reviews the computation of the grand canonical potential of neutron matter in a magnetic field.  As this  has been discussed extensively elsewhere in the literature, the review will be brief and only serves to summarise the main results and notation.  Section \ref{two} develops a way of handling the divergencies and discusses the unambiguous computation of the full partition function, which includes the vacuum energy contribution,  for neutron matter in a magnetic field.  Section \ref{three} presents the results of these computations and discusses the effects of the vacuum energy contribution.  We emphasise that pure neutron matter is probably not a good description of a neutron star and that protons, leptons and mesons should be included as was done in \cite{Lattimer2, Diener3}.   However, the aim of this paper is not a full quantitative analysis of the neutron star equation of state with the vacuum energy included, but rather an attempt to identify the possible physical effects that the vacuum energy may have when computed in a controlled way and in the simplest possible setting.   Finally, Section \ref{four} summarises and draws conclusions.

\section{Grand canonical potential of pure neutron matter}
\label{one}
Pure neutron matter coupled to an electromagnetic field is described by the following action ($\hbar=c=1$) \cite{Lattimer2,Diener3}
\begin{equation}
S\left[A^\mu,\bar\psi,\psi\right]=\int d^4x\left[\bar{\psi}{ (x)}\left[i\gamma^{\mu}\partial_{\mu}-\frac{g}{2}F^{\mu\nu}\sigma_{\mu\nu}-m\right]\psi{ (x)}-\frac{1}{4}F^{\mu\nu}F_{\mu\nu}\right],\label{Lag}
\end{equation}
where  $m$ is the neutron mass, $F^{\mu\nu}=\partial^\mu A^\nu-\partial^\nu A^\mu$, and $\sigma^{\mu\nu} = \frac{i}{2}\left[\gamma^\mu,\gamma^\nu\right]$ are the generators of the Lorentz group. We use Heaviside-Lorentz units, the convention of \cite{itzykson} for the $\gamma$-matrices and metric $g^{00}=1$, $g^{ii}=-1$. We have also included an anomalous coupling term with strength $g$, which takes care of the coupling between the anomalous magnetic dipole moment of a neutron and the electromagnetic field in a manifestly Lorentz covariant way. Note that the anomalous magnetic dipole moment of the neutron arises from its composite nature and is therefore inherently different to the anomalous magnetic dipole moment of an electron, which is a quantum effect.  As the neutron is electrically neutral, this is the only coupling with the electromagnetic field.  The coupling strength $g$ is dimensionful (with the units of fm, i.e. an inverse mass), which renders the theory non-renormalizable and creates a challenge in terms of the unambiguous computation of the contributions arising from quantum fluctuations.  The last term in (\ref{Lag}) is the standard Lagrangian for the electromagnetic field.  Finally, note that this Lagrangian is still manifestly gauge invariant (keep in mind that the neutron fields are not transformed as they are electrically neutral).

We want to compute the grand canonical partition function and potential at finite temperature and density.  To do this we first have to transform to Euclidean space in the usual way by setting $x^0_{\rm E}=ix^0=ix_0$, $x^i_{\rm E}=x^i=-x_i$.  We also introduce Euclidean Gamma matrices with the properties $\left\{\gamma^\mu_{\rm E},\gamma^\nu_{\rm E}\right\}=2\delta^{\mu\nu}$ and $\left(\gamma^\mu_{\rm E}\right)^\dagger=\gamma^\mu_{\rm }$.  Now $\sigma^{\mu\nu}_{\rm E} = \frac{i}{2}\left[\gamma^\mu_{\rm E},\gamma^\nu_{\rm E}\right]$ are the generators of $SO(4)$.  The Euclidean action, which includes the chemical potential $\mu$, reads
\begin{equation}
S_{\rm E}\left[A^\mu_{\rm E},\bar\psi,\psi\right]=\int d^4x_{\rm E}\left[\bar{\psi}{ (x)}\left[\gamma^{\mu}_{\rm E}\partial^{\mu}_{\rm E}+\frac{g}{2}F^{\mu\nu}_{\rm E}\sigma^{\mu\nu}_{\rm E}+m+\mu\gamma^0_{\rm E}\right]\psi{ (x)}+\frac{1}{4}F^{\mu\nu}_{\rm E}F^{\mu\nu}_{\rm E}\right].\label{LagE}
\end{equation}
Here the Euclidean time integration is over the interval $[0,\beta]$ with $\beta$ the inverse temperature.  The grand canonical partition function is obtained from the functional integral
\begin{equation}
\label{parff}
Z[J]=\int\left[dA^\mu_{\rm E}\right]\left[d\bar\psi\right]\left[d\psi\right]e^{-S_{\rm E}\left[A^\mu,\bar\psi,\psi\right]+\int d^4x_{\rm E}J^\mu_{\rm E} A^\mu_{\rm E}},
\end{equation}
where the integral is taken over bosonic and fermionic field configurations that are respectively periodic and anti-periodic in the Euclidean time direction.  We have also included sources for the gauge potentials.  For gauge invariance, we must require $\partial^\mu_{\rm E}J^\mu_{\rm E}=0$.  These sources represent external physical currents to which the electromagnetic field couples.  In what follows we drop the subscript E referring to Euclidean as this will be clear from the context.  

Our first step is to integrate out the fermionic fields, which gives for the partition function
\begin{equation}
\label{parf}
Z[J]=\int\left[dA^\mu\right]e^{-S_{\rm eff}\left[A^\mu\right]+\int d^4x J^\mu A^\mu},
\end{equation}
with the effective action
\begin{equation}
S_{\rm eff}\left[A^\mu\right]=\int d^4x\left[\frac{1}{4}F^{\mu\nu}F_{\mu\nu}\right]-{\rm Tr}\log\left[\gamma^{\mu}\partial^{\mu}+\frac{g}{2}F^{\mu\nu}\sigma^{\mu\nu}+m+\mu\gamma^0\right].\label{Lageff}
\end{equation}

The simplest way to proceed is to compute (\ref{parf}) in a saddle point approximation.  Introducing the classical field configuration $\bar{A}\left[J\right]^\mu$ that solves the classical equation of motion
\begin{equation}
\frac{S_{\rm eff}\left[A^\mu\right]}{\delta A^\mu\left(x\right)}=J^\mu\left(x\right),
\end{equation}
one shifts the integration variables in (\ref{parf}) to $A^\mu=\bar{A}^\mu+a^\mu$, expands (\ref{Lageff}) to second order in the fluctuations $a^\mu$ and performs the resulting Gauss integral.  Doing this one notices that there is no longer an explicit dependence of $Z$ on the sources, but that this dependence now enters via the dependence of $\bar{A}\left[J\right]^\mu$ on the sources.  It is therefore convenient to now rather adopt the classical field configurations, rather than the sources, as the variables and to write
\begin{equation}
Z\left[\bar{A}^\mu\right]\equiv e^{-Q\left[\bar{A}^\mu\right]}=e^{-S_{\rm eff}\left[\bar{A}^\mu\right]-\frac{1}{2}{\rm Tr}\log\left[\frac{\delta^2 S_{\rm eff}}{\delta^\mu\left(x\right)A^\nu\left(y\right)}\right]_{A^\mu=\bar{A}^\mu}},
\end{equation}
where we have introduced the grand canonical potential
\begin{equation}
Q\left[\bar{A}^\mu\right]=S_{\rm eff}\left[\bar{A}^\mu\right]+\frac{1}{2}{\rm Tr}\log\left[\frac{\delta^2 S_{\rm eff}}{\delta^\mu\left(x\right)A^\nu\left(y\right)}\right]_{A^\mu=\bar{A}^\mu}.
\end{equation}

The one loop corrections to the effective action are divergent, but these divergencies cannot be absorbed in the traditional way as the theory is non-renormalisable.  It is therefore customary to truncate at the tree level in which case the grand canonical potential is given by
\begin{equation}
Q\left[\bar{A}^\mu\right]=S_{\rm eff}\left[\bar{A}^\mu\right]=\int d^4x\left[\frac{1}{4}\bar{F}^{\mu\nu}\bar{F}_{\mu\nu}\right]-{\rm Tr}\log\left[\gamma^{\mu}\partial^{\mu}+\frac{g}{2}\bar{F}^{\mu\nu}\sigma^{\mu\nu}+m+\mu\gamma^0\right].\label{Qpot}
\end{equation}

We remark that it is more conventional to carry out the discussion above using average fields and the effective action related to $W\left[J\right]$ through a Legendre transformation.  Doing this, one ends up with the same result at the one-loop level, but the discussion above has the added advantage of emphasising the physical meaning of the background fields.  

Regarding the higher order corrections, we would like to mention that there are non-perturbative ways of treating such effective theories using the so called exact renormalisation group equations (ERGE) \cite{wetter}.  Several scalar and 4-fermion effective theories have been treated successfully in this way \cite{wetter}, but here this fails due to fact that the regularisation breaks the gauge invariance.  Several attempts to treat gauge theories in this way have been made in e.g. \cite{wetter1,bon}, but a satisfactory treatment is still lacking.  

It is not possible to compute (\ref{Qpot}) for arbitrary field configurations.  However, if we assume the sources (currents) $J^\mu$ are stationery and homogenous, at least on some macroscopic scale, the classical field configurations $\bar{A}^\mu$ will inherit this feature.   This implies that there are no electric fields and that the magnetic field is a spatially constant background field, which we can take without loss of generality to be pointing in the $z$-direction. Based on these assumptions we choose the gauge field to be $A^\mu=(0,0,Bx,0)$, where the magnetic field, $B$, has the units fm$^{-2}$.  This choice simplifies  $\frac{1}{4}\bar{F}^{\mu\nu}\bar{F}_{\mu\nu}$ in (\ref{Qpot}) to $\frac{1}{2} B^2$ and reduces the anomalous coupling to
\begin{equation}
		\frac{g}{2}F^{\mu\nu}\sigma_{\mu\nu} =- g{\bm\Sigma}\cdot {\bm B},
\end{equation}
where ${\bm\Sigma} = {\bm \sigma}\otimes {\bm 1_2}$. Here ${\bm \sigma}$ are the Pauli matrices, and ${\bm 1_2}$ is the $2\times 2$ identity matrix. 

The steps to compute the grand canonical potential from here are well documented in the literature \cite{kap}.  Introducing $\Omega=\frac{Q}{\beta V}$, with $V$ the volume, one has \begin{eqnarray}
	\Omega=&&- \sum_\lambda\int\frac{dk_z dk_\perp k_\perp}{(2\pi)^2}\left[\omega\left(k_z,k_\perp,\lambda\right)+\frac{1}{\beta}\log\left(1+e^{-\beta(\omega\left(k_z,k_\perp,\lambda\right)-\mu)}\right)+\frac{1}{\beta}\log\left(1+e^{-\beta(\omega\left(k_z,k_\perp,\lambda\right)+\mu)}\right)\right]\nonumber\\
	&&+\frac{1}{2}B^2.\label{gcanp}
\end{eqnarray}
Here we have introduced the momentum perpendicular to the magnetic field, $k_\perp$ [fm$^{-1}$], and the momentum parallel to the magnetic field $k_z$ [fm$^{-1}$], so that $k=\sqrt{k_\perp^2+k_z^2}$. In terms of these the positive single particle energies with units fm$^{-1}$ are given by
\begin{equation}
\omega\left(k_z,k_\perp,\lambda\right)= \sqrt{\left(\sqrt{k_{\bot}^2+{m}^2}+\lambda g B\right)^2+k_{z}^2}, \label{spe}
\end{equation}
where $\lambda=\pm 1$ labels the spin in the z-direction.  
Note that $\Omega$ is dimensionful with dimension fm$^{-4}$.

Once $\Omega$ is known, the computation of  other thermodynamic quantities is straightforward.  The quantities most relevant to our present discussion are the particle density, $\rho$, and the energy density, $\epsilon$, given by 
\begin{eqnarray}
\rho&=&-\frac{\partial\Omega}{\partial\mu},\label{pdens}\\
\epsilon&=&\frac{\partial\left(\beta\Omega\right)}{\partial\beta}.\label{endens}
\end{eqnarray}
Note that $\rho$ has the dimensions fm$^{-3}$ and $\epsilon$ the dimensions fm$^{-4}$.

In the zero temperature limit ($\beta\rightarrow\infty$) (\ref{gcanp}) reduces to
\begin{equation}
	\Omega=\sum_\lambda\int\frac{d^3k}{(2\pi)^3}\big[-\omega\left(k_z,k_\perp,\lambda\right)+\left(\omega\left(k_z,k_\perp,\lambda\right)-\mu\right)\Theta\left[\mu-\omega\left(k_z,k_\perp,\lambda\right)\right]\big]+\frac{1}{2}B^2,\label{gcanzt}
\end{equation}
and the energy density is given by 
\begin{equation}
	\epsilon=\sum_\lambda\int\frac{d^3k}{(2\pi)^3}\big[-\omega\left(k_z,k_\perp,\lambda\right)+\omega\left(k_z,k_\perp,\lambda\right)\Theta\left[\mu-\omega\left(k_z,k_\perp,\lambda\right)\right]\big]+\frac{1}{2}B^2.\label{edens}
\end{equation}

Another important thermodynamic quantity is the magnetisation, $\vec{M}$, defined as the dipole moment per unit volume, and associated with it the applied magnetic field $\vec{H}$.  As we take the magnetic field $\vec{B}$ to be in the z-direction, we also assume these quantities to be in the z-direction and ignore their vectorial properties in what follows.  The first point to emphasise is that the magnetic field $B$ is the quantity that we observe and which determines the vector potential $A^\mu$ and therefore also the quantity that couples to the neutron magnetic dipole moment \cite{Bland} as indicated in (\ref{LagE}).  However, we cannot control $B$ due to presence of magnetic dipoles that also contribute to $B$. The quantity we can control through appropriate current configurations is the applied magnetic field $H$, defined by $H=B-M$.  Note that in our units $H$, $B$ and $M$ have the same dimensions (fm$^{-2}$) and that there are no factors of $4\pi$.  In pure neutron matter there are no free charges and we expect $H=0$, but more generally the presence of charged particles, such as protons, will result in $H\ne 0$.  We therefore treat $H$ as a free parameter in what follows.  

In a magnetized system the magnetisation contributes a term $-MB$ to the system's energy density.  If we include the contribution of the magnetic field in the energy density, i.e the term $\frac{B^2}{2}$ as is done in (\ref{endens}), the change in the energy density due to a change in the magnetic field at fixed magnetisation is 

\begin{equation}
\Delta\epsilon=B\Delta B-M\Delta B=H\Delta B
\end{equation}
and we conclude \cite{landau,Lattimer2}
\begin{equation}
\label{H}
H=\frac{\partial\epsilon}{\partial B}.
\end{equation}
As was elaborated in \cite{Lee} this should actually be seen as a self-consistency condition for $M$, i.e. given $H$, we must determine $M$ from this equation (keep in mind that $B=H+M$).  In particular, note that for $H=0$, this is a minimisation condition for the energy density.  Of course (\ref{H}) can be written in many different ways using standard thermodynamic identities, but we find this to be the most convenient for our present purposes.

One can also define the longitudinal and transverse pressures \cite{Ferr4} 
\begin{eqnarray}
	P_{\parallel} = -\Omega,\  P_{\perp} = -\Omega+B\frac{\partial\Omega}{\partial B}.\label{presb}
\end{eqnarray}

It should be clear that (\ref{gcanp}) is a divergent expression due to the first term that presents the vacuum energy contribution to $\Omega$.  From (\ref{endens}) one notes that the same term also appears in $\epsilon$ and that this is subsequently also a divergent expression.   On the other hand, since the single particle energies do not depend on $\mu$, this term drops out in the particle density (\ref{pdens}) so that this is a finite expression.  One way of handling this divergency is to simply introduce a momentum cut-off, such as was done in \cite{Ferr4}, but doing this introduces an explicit dependence of $\Omega$ on the cut-off, which is introduced in a rather ad hoc way.  One must then find physical grounds for a reasonable choice of this cut-off.  In \cite{Ferr3} this was done by taking the cut-off to be of the order of the chemical potential, which seems physically reasonable, but is still somewhat ad hoc. 

In the next section we present an alternative way of handling this divergency, which is much more controlled and systematic and yields an unambiguous result for the vacuum energy contribution.

\section{Computing the vacuum energy contribution to the grand canonical potential}
\label{two}

It is convenient to rewrite (\ref{gcanp}) in the following way
\begin{equation}
\label{dlgcanp}
\Omega\left(m,g,B,\mu,\beta\right)=\Omega_{\rm Vac}\left(m,gB\right)+\Omega_{\rm N}\left(m,gB,\mu,\beta\right)+\Omega_{\rm B}\left(B\right)
\end{equation}
where
\begin{eqnarray}
\Omega_{\rm Vac}\left(m,gB\right)=&&- \sum_\lambda\int\frac{dk_z dk_\perp k_\perp}{(2\pi)^2}\omega\left(k_z,k_\perp,\lambda\right),\label{vac}\\
\Omega_{\rm N}\left(m,gB,\mu,\beta\right)=&&- \sum_\lambda\int\frac{dk_z dk_\perp k_\perp}{(2\pi)^2}\left[\frac{1}{\beta}\log\left(1+e^{-\beta(\omega\left(k_z,k_\perp,\lambda\right)-\mu)}\right)+\frac{1}{\beta}\log\left(1+e^{-\beta(\omega\left(k_z,k_\perp,\lambda\right)+\mu)}\right)\right],\nonumber\\
\label{Neutron}\\
\Omega_{\rm B}\left(B\right)=\frac{1}{2}B^2.\label{Maxwell}
\end{eqnarray}

The only divergent quantity here is $\Omega_{\rm Vac}$ and for the moment we focus on it.  From a simple scaling argument we note from (\ref{vac}) that 
\begin{equation}
\Omega_{\rm Vac}\left(m,gB\right)=m^4\Omega_{\rm Vac}\left(1,\frac{gB}{m}\right)\equiv m^4 f\left(x\right),\label{scale}
 \end{equation}
where we have introduced the dimensionless quantity $x=\frac{gB}{m}$ and 
\begin{equation}
f\left(x\right)=-\int\frac{dq_z dq_\perp q_\perp}{(2\pi)^2}h\left(q_\perp,q_z,x\right) \label{f}
\end{equation}
with
\begin{equation}
h\left(q_\perp,q_z,x\right)=\sqrt{\left(\sqrt{q_{\perp}^2+1}+x\right)^2+q_{z}^2}+\sqrt{\left(\sqrt{q_{\perp}^2+1}-x\right)^2+q_{z}^2}.\label{fint}
\end{equation}
Note that the integration variables $q$ are dimensionless here.  

It is convenient to rather work in spherical coordinates by setting $q_{\perp}=q\sin\theta$, $q_z=q\cos\theta$ where $q\in[0,\infty)$ and $\theta\in[0,\pi]$ in which case (\ref{f}) reads
\begin{equation}
f\left(x\right)=-\int\frac{dq q^2}{(2\pi)^2}h\left(q,\theta,x\right) . \label{fs}
\end{equation}
We can now analyse the divergent behaviour of $f\left(x\right)$ and its derivatives by performing an asymptotic expansion of the integrand in (\ref{fs}) in $q$.  From this we note that $f\left(x\right)$ has a quartic divergence.   The first and second derivatives are quadratically divergent.  The third and fourth derivatives are logarithmic divergent, but the fifth and higher order derivatives are finite.  We therefore conclude that $f\left(x\right)$ must satisfy the unambiguous and finite differential equation
\begin{equation}
\frac{d^5 f\left(x\right)}{dx^5}=-\int\frac{dq_z dq_\perp q_\perp}{(2\pi)^2}\frac{\partial^5 h\left(q_\perp,q_z,x\right) }{\partial x^5}.\label{DE}
\end{equation}
Upon integrating this equation, we have to specify five dimensionless integration constants.  On this level, this is where the ambiguity in the original undefined expression (\ref{vac}) resides.  We only obtain an unambiguous result once these integration constants have been fixed in some way.  We therefore conclude that 
\begin{equation}
\label{Omren}
\Omega\left(m,g,B,\mu,\beta\right)=m^4\left[\sum_{n=0}^4\frac{c_n}{n!}\left(\frac{gB}{m}\right)^n+f_0\left(\frac{gB}{m}\right)\right]+\Omega_{\rm N}\left(m,gB,\mu,\beta\right)+\Omega_{\rm B}\left(B\right).
\end{equation}
Here $f_0\left(x\right)$ is a solution of (\ref{DE}) with initial conditions 
\begin{equation}
\left(\frac{d^n f_0\left(x\right)}{dx^n}\right)_{x=0}=0,\;\forall n=0,1,2,3,4,
\end{equation}
and $c_n$ are dimensionless integration constants.  

To proceed we compute the function $f_0\left(x\right)$ explicitly.  We find
\begin{equation}
\label{vacc}
f_0\left(x\right)=\frac{-13 \left(x^2-6\right) x^2+6 \left(x^4-6 x^2-3\right) \log \left(1-x^2\right)-96 x \tanh ^{-1}(x)}{288 \pi ^2}.
\end{equation}
We note that this expression is singular at $x=\frac{gB}{m}=\pm 1$. At and above this threshold the energy gap between the different orientations of the neutron dipole moment, $\lambda = \pm1$ in (\ref{spe}), exceeds $2m$, which makes spontaneous pair generation possible and leads to an instability.   

To fix the constants $c_n$, we return to (\ref{LagE}) and (\ref{parff}) and note that in the $m\rightarrow\infty$ limit the fermion functional integral is localised on the classical configurations $\bar\psi=\psi=0$.  In this limit the grand canonical potential must therefore coincide, up to a possible divergent magnetic field independent constant, with the classical energy density of a constant magnetic field $\Omega=\frac{1}{2}B^2$.   From (\ref{vacc}),  we verify $\lim_{m\rightarrow\infty}m^4f_0\left(\frac{gB}{m}\right)=0$.  Since $\Omega_N$ vanishes in this limit, we note from (\ref{Omren}) that $c_n=0$, $\forall n\ne 0$.  The coefficient $c_0$ must be chosen to cancel any divergent contributions and we find the final result
\begin{equation}
\label{Omrenf}
\Omega\left(m,g,B,\mu,\beta\right)=m^4f_0\left(\frac{gB}{m}\right)+\Omega_{\rm N}\left(m,gB,\mu,\beta\right)+\Omega_{\rm B}\left(B\right),
\end{equation}
with $f_0\left(x\right)$ given in (\ref{vacc}).

\section{Results}
\label{three}

In the rest of this section we adopt the following values for the parameters: We take the mass, $m$, and anomalous coupling, $g$, to be the values of a free neutron  i.e. $m=4.761$ fm$^{-1}$, $g=0.031$ fm.  We also note that the conversion factor for the magnetic field in units of fm$^{-2}$ and gauss is 1 fm$^{-2}=1.993\times 10^{18}$ G.  This implies that for any realistic values of the magnetic field, the dimensionless parameter $x$ is rather small and of the order $10^{-3}$.  

Where temperature effects are considered, $\beta$ is either considered at the zero temperature limit ($\beta\rightarrow\infty$) or at an arbitrary value of $\beta=20$ fm, which corresponds to an energy of about 1\% of $m$. 

\begin{figure}[t]
    \centering
   \includegraphics[width=0.5\textwidth]{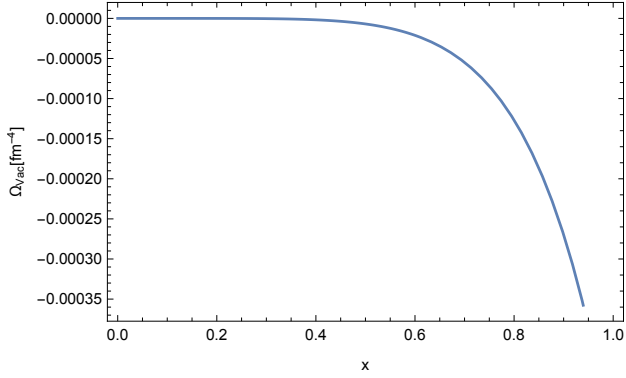}
  \caption{The unambiguous vacuum contribution to $\Omega$ as a function of $x=\frac{gB}{m}$.  We note that it is a very small contribution at typical values of $x\approx 10^{-3}$. }
  \label{vacun}
\end{figure}

Let us start by computing the unambiguous vacuum contribution $\Omega_{\rm Vac}$ to $\Omega$.  Figure \ref{vacun} shows this quantity as a function of $x$.  We note that for any realistic values of the magnetic field this a very small contribution to $\Omega$.     

\begin{figure}[t]
    \centering
   \begin{tabular}{c c}
    (a)&(b)\\
  \includegraphics[width=0.5\textwidth]{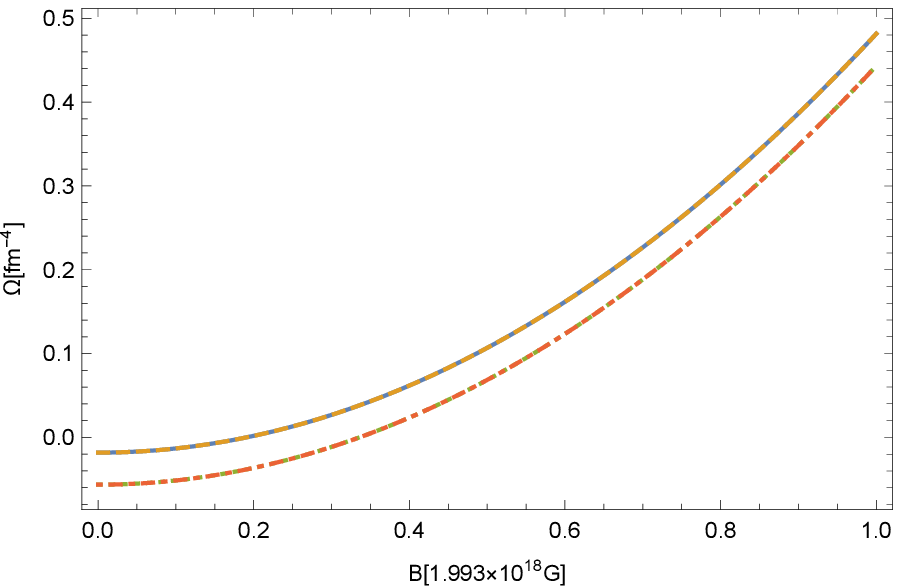}&\includegraphics[width=0.5\textwidth]{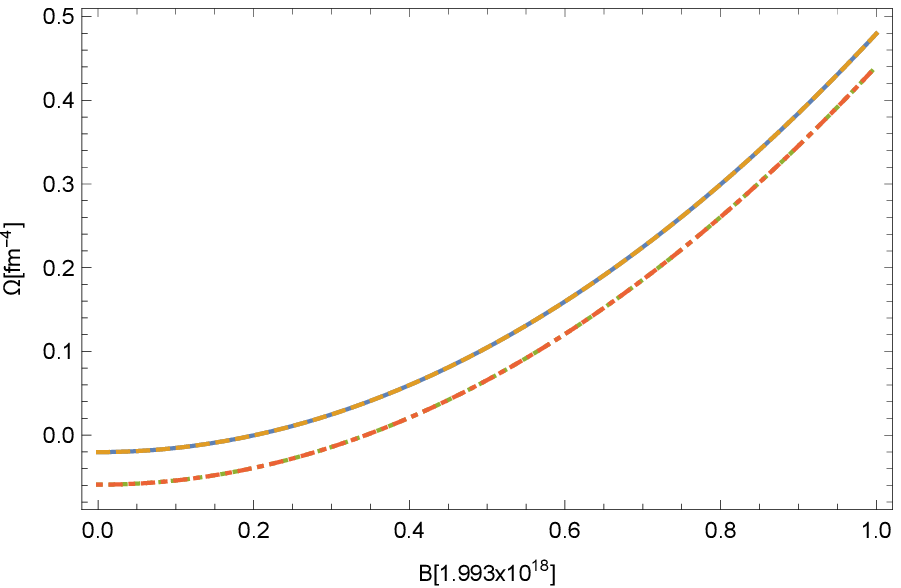}
	\end{tabular}
  \caption{(a) $\Omega$ at zero temperature with and without vacuum corrections and at $\rho=\rho_s$ (solid and dashed  lines) and $\rho=2\rho_s$ (dashed and dot-dashed lines).  (b) Same as in (a), but now for $\beta=20$ fm.   The vacuum corrections are not discernable on these graphs as they are very small}
  \label{OmE}
\end{figure}

Figure \ref{OmE} shows $\Omega$ with and without the vacuum energy contribution at (a) zero temperature and (b) $\beta=20$ fm and two densities $\rho=\rho_s$ and $\rho=2\rho_s$ where $\rho_s=0.16$ fm$^{-3}$ is saturation density.

\begin{figure}[t]
    \centering
   \begin{tabular}{c c}
    (a)&(b)\\
  \includegraphics[width=0.5\textwidth]{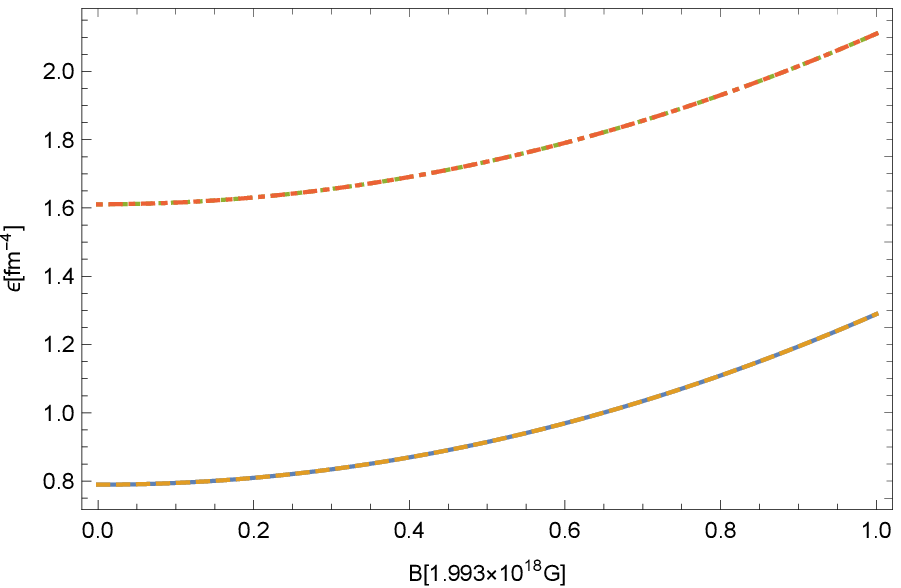}&\includegraphics[width=0.5\textwidth]{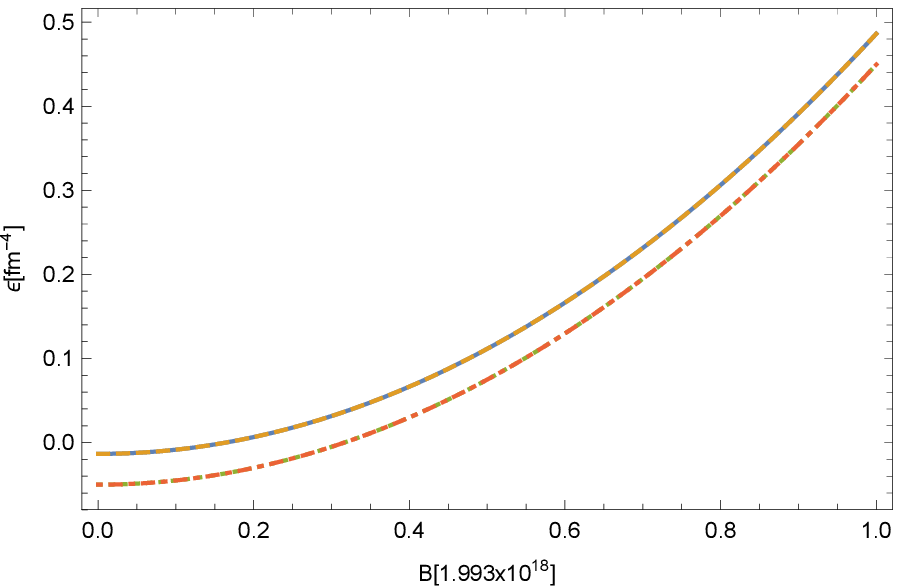}
	\end{tabular}
  \caption{(a) $\epsilon$ at zero temperature with and without vacuum corrections and at $\rho=\rho_s$ (solid and dashed  lines) and $\rho=2\rho_s$ (dashed and dot-dashed lines).  (b) Same as in (a), but now for $\beta=20$ fm.   The vacuum corrections are not discernable on these graphs as they are very small}
  \label{EnE}
\end{figure}

Figure \ref{EnE} shows $\epsilon$ with and without the vacuum energy contribution at (a) zero temperature and (b) $\beta=20$ fm and two densities $\rho=\rho_s$ and $\rho=2\rho_s$.

\begin{figure}[t]
    \centering
   \begin{tabular}{c c}
    (a)&(b)\\
  \includegraphics[width=0.5\textwidth]{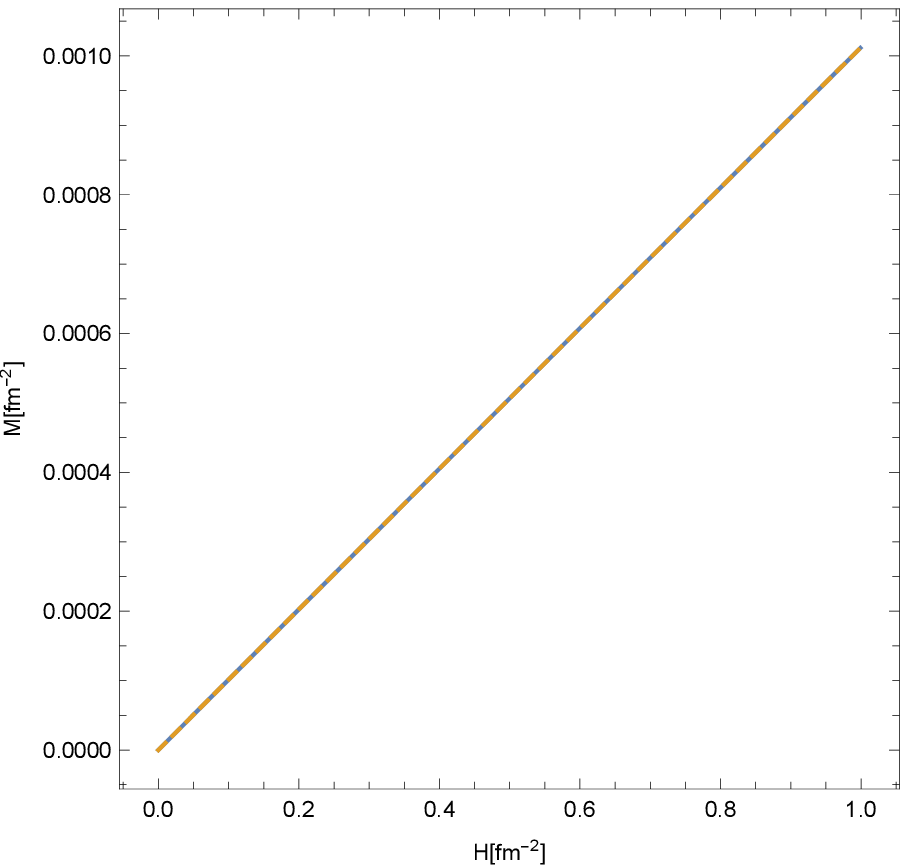}&\includegraphics[width=0.5\textwidth]{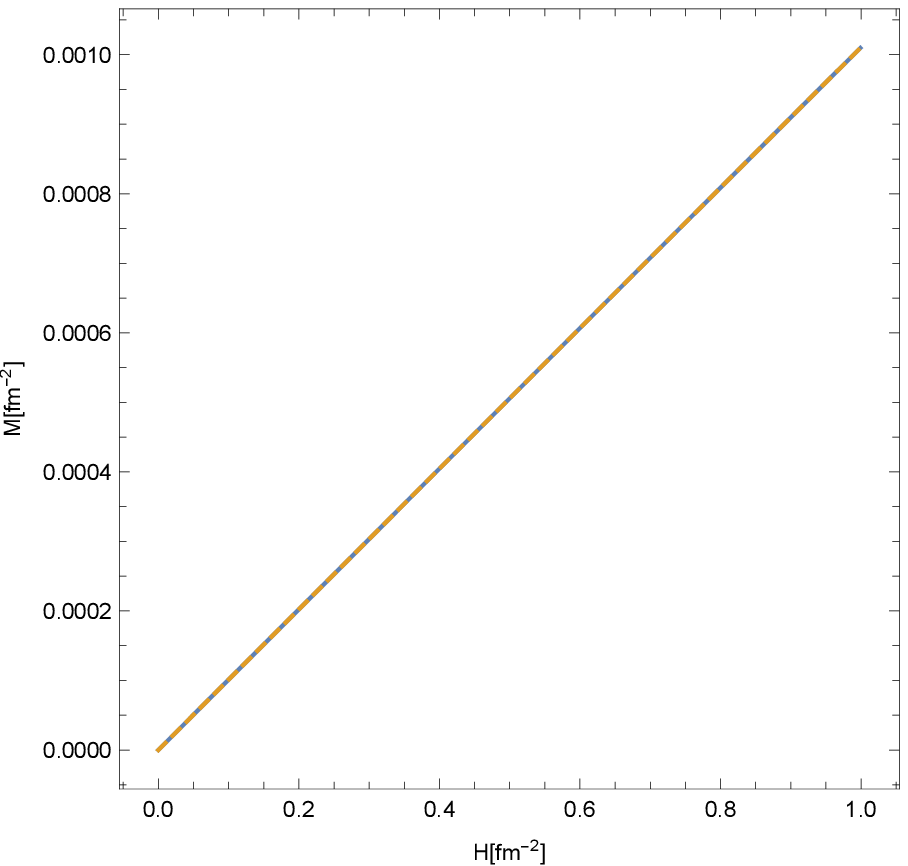}
	\end{tabular}
    \caption{$M$ as a function of the applied magnetic field $H$ at (a) zero temperature and $\rho=2\rho_s$ with (dashed line) and without (solid line) the  vacuum energy contribution.  (b) Same as in (a), but now for $\beta=20$ fm.}
  \label{MvH}
\end{figure}

Figure \ref{MvH} shows $M$ as a function of $H$ at (a) zero temperature and (b) $\beta=20$ fm at $\rho=2\rho_s$ with and without the vacuum energy contribution.  As before the vacuum contribution is very small and cannot be discerned on the graph.  We also note a strict paramagnetic behaviour with $M$ vanishing at vanishing $H$.

\begin{figure}[t]
    \centering
  \begin{tabular}{c c}
    (a)&(b)\\
  \includegraphics[width=0.5\textwidth]{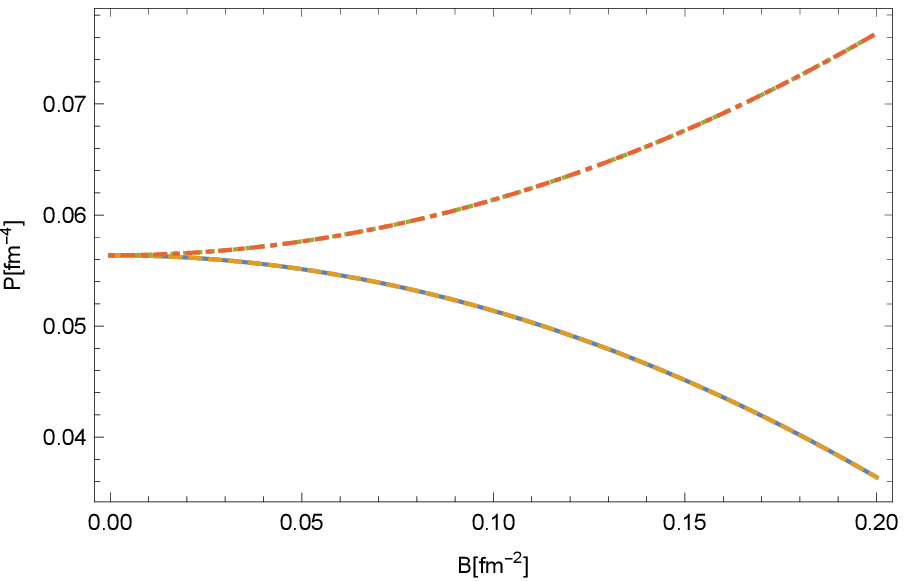}&\includegraphics[width=0.5\textwidth]{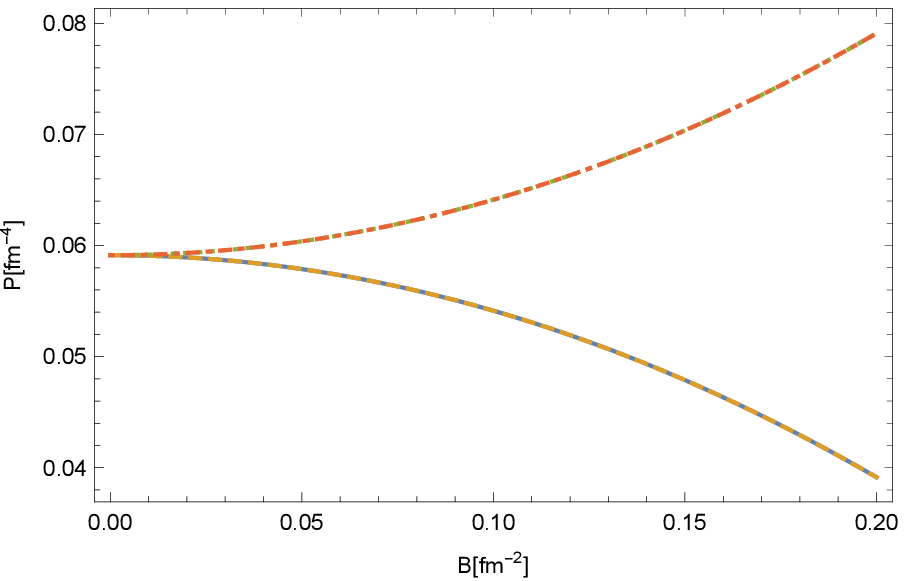}
	\end{tabular}
  \caption{Longitudinal pressure without (solid line) and with (dashed line) the vacuum energy contribution as a function of the magnetic field.  Also shown is the transverse pressure without (dotted line) and with (dot-dashed line) the vacuum energy contribution.  (a) At zero temperature and (b) at $\beta=20$ fm with constant density $\rho=2\rho_s$.  }
  \label{press}
\end{figure}

Finally, Figure \ref{press} shows the longitudinal and transverse pressures as a function of the magnetic field at constant density $\rho=2\rho_s$ and (a) zero temperature and (b) $\beta=20$ fm.  The conclusions are as before.

\section{Conclusion}
\label{four}

Theories with anomalous couplings play an important role in the description of nuclear matter in magnetic fields and are as such very important in applications to neutron stars where neutrons dominate and can only couple anomalously to the strong magnetic field.  These theories are, however, non-renormalisable which renders a perturbative renormalisation approach ineffective.   One way to escape this dilemma is to argue that it is an effective theory that cannot be trusted beyond some momentum scale and to introduce some momentum cut-off.  This introduces a dependence of the physical quantities on the cut-off scale, which may be chosen rather ad hoc, unless one has convincing physical grounds for the choice of cut-off scale.  Here we have demonstrated that, at least as far as the vacuum energy is concerned, this can be avoided and the vacuum energy contribution can be computed unambiguously.  For neutron matter in a magnetic field we found that the vacuum energy contribution is generally very small and does not substantially affect any of the observables.  

It is not clear how the higher order loop corrections may affect these results.  As already mentioned, they cannot be treated in a perturbative approach as the theory is non-renormalisable.  Taking the point of view of effective theories, one may hope to treat it using non-perturbative ERGE techniques \cite{wetter}. Unfortunately, this program also runs into difficulties due to the breaking of gauge invariance by the regularisation used in this approach.  It is therefore not clear if the effective potential may develop non-trivial features when higher order corrections are included and whether a possible ferromagnetic phase transition may occur as was suggested in \cite{Lee} for charged particles.  At the current level of approximation this is certainly not possible.

\section{Acknowledgements}

This research is supported by the National Research Foundation of South Africa. FGS also acknowledges generous support from BIUST where part of this work was completed.

\end{document}